\documentclass{article}

\usepackage[final,nonatbib]{neurips_2020_ml4ps}

\usepackage[utf8]{inputenc} 
\usepackage[T1]{fontenc}    
\usepackage{hyperref}       
\usepackage{url}            
\usepackage{booktabs}       
\usepackage{amsfonts}       
\usepackage{nicefrac}       
\usepackage{microtype}      
\usepackage[normalem]{ulem}
\usepackage{graphicx}       
\usepackage{subfig}         
\usepackage[dvipsnames]{xcolor} 
\usepackage{gensymb}        
\usepackage{cite}
\usepackage{upgreek}
\iffalse

\newcommand{\nsection}[1]{\section{#1}}

\else

\newcommand{\nsection}[1]{\vspace{-0.1cm}\section{#1}\vspace{-0.1cm}}

\fi

\title{Deep-Learning-Based Kinematic Reconstruction for DUNE}

\author{
  \parbox[c]{0.85\linewidth}{\centering
    Junze Liu$\mathrm{^{1}}$, Jordan Ott$\mathrm{^1}$, Julian Collado$\mathrm{^1}$, Benjamin Jargowsky$\mathrm{^1}$, Wenjie Wu$\mathrm{^1}$, Jianming Bian$\mathrm{^1}$, Pierre Baldi$\mathrm{^2}$}\vspace{1mm}\\
  (For the DUNE Collaboration)\vspace{1mm}\\
  University of California, Irvine\\
  Irvine, CA 92617 \\
  \parbox[c]{0.88\linewidth}{\centering 
    \texttt{\{junzel1, jott1, colladou, bjargows, wenjieww, bianjm\}@uci.edu}$\mathrm{^1}$} \\
  \texttt{pfbaldi@ics.uci.edu}$\mathrm{^2}$
}

\begin{document}

\maketitle

\begin{abstract}
\label{sec:abstract}
  In the framework of three-active-neutrino mixing, the charge parity phase, the neutrino mass ordering, and the octant of $\theta_{23}$ remain unknown. The Deep Underground Neutrino Experiment (DUNE) is a next-generation long-baseline neutrino oscillation experiment, which aims to address these questions by measuring the oscillation patterns of $\nu_\mu/\nu_e$ and $\bar\nu_\mu/\bar\nu_e$ over a range of energies spanning the first and second oscillation maxima. DUNE far detector modules are based on liquid argon TPC (LArTPC) technology. A LArTPC offers excellent spatial resolution, high neutrino detection efficiency, and superb background rejection, while reconstruction in LArTPC is challenging. Deep learning methods, in particular, Convolutional Neural Networks (CNNs), have demonstrated success in classification problems such as particle identification in DUNE and other neutrino experiments. However, reconstruction of neutrino energy and final state particle momenta with deep learning methods is yet to be developed for a full AI-based reconstruction chain. To precisely reconstruct these kinematic characteristics of detected interactions at DUNE, we have developed and will present two CNN-based methods, 2-D and 3-D, for the reconstruction of final state particle direction and energy, as well as neutrino energy. Combining particle masses with the kinetic energy and the direction reconstructed by our work, the four-momentum of final state particles can be obtained. Our models show considerable improvements compared to the traditional methods for both scenarios.
  
\end{abstract}
\vspace{-0.1cm}
\nsection{Introduction}
\label{sec:introduction}
\vspace{-0.1cm}
Neutrino oscillation is the first experimental observation beyond the standard reconstruction method which provides evidence of neutrinos with a non-zero mass. This phenomenon originates from the mixture between the mass and flavor eigenstates of neutrinos, and is commonly described by the PMNS formalism with six fundamental parameters~\cite{Pontecorvo:1957cp, Pontecorvo:1967fh, Maki:1962mu}. 
DUNE aims to make precise measurements of these oscillation parameters through the detection of $\nu_\mu$/$\bar\nu_\mu$ disappearance and $\nu_e$/$\bar\nu_e$ appearance over a long propagation distance~\cite{Abi:2020evt}. A precise neutrino energy reconstruction provides a chance to estimate the neutrino oscillation parameters with a high significance. Neutrinos are normally detected via charged-current (CC) interactions with the nuclei in the detector. In a CC interaction, the final state includes a charged lepton with the same flavor of the incident neutrino, which in the case of DUNE it is either an electron or a muon, and one or more hadrons. Directions and energies (momentum) of these final state particles give the full kinematics of an neutrino interaction.

Traditionally, the energies of electrons and hadrons are calculated from calorimetric energies and calibration factors.  The kinetic energy of the muons is reconstructed using the length of the track or the multiple Coulomb scattering method~\cite{Abratenko:2017nki}, depending on whether the track is contained or not inside the detector. The Coulomb scattering method uses the average scattering angle of a muon to predict the energy not contained in the detector. Directions of particles are reconstructed by fitting to detector hits. The neutrino energy is obtained as the sum of the lepton and hadron energies \cite{TraditionalMethod}.
The reconstruction of these kinematic parameters is challenging in DUNE due to missing energy caused by argon impurities, nonlinear detector energy responses and overlapping particle trajectories.
Image recognition models like convolutional neural networks (CNNs)~\cite{lecun1989backpropagation,baldi93finger}  have demonstrated outstanding performance in classification tasks using calorimeter images at DUNE and other high energy physics experiments \cite{Abi:2020DUNEDL, Baldi:NatureHEP, DLHEP:IgorOstrovsky, Almeida:2015jua, de_Oliveira_jet_images_deep_learning_edition_2016, Komiske:2016DLInColor, Guest:JetFlavorClassification, Sadowski:DarkKnowledge,baldi2020deep,belayneh2020calorimetry}. Nevertheless, applications of CNNs to solve regression problems and reconstruct continuous variables in neutrino physics is still preliminary~\cite{ NOvABaldi, Delaquis:2018zqi}, especially when the variables are vectors. In this work, we propose using CNNs to reconstruct the aforementioned kinematic parameters directly from DUNE's 2-D and 3-D LArTPC images. Our work demonstrates that the kinematics of a physics process in a complicated detector can be fully reconstructed by AI without laborious human-engineered algorithms.

To detect the interactions, the DUNE LArTPC far detector has 3 wire planes for readout, positioned at different angles from each other. After a number of corrections, the signals read out by these planes are reconstructed as "hits", categorized by the charge per hit, the wire and wire plane in which the hit occurred on, and the time of the hit, in units of ticks, which are 0.5 $\upmu$s each. 
This information can be represented as 3 images for each neutrino event, one for each wire plane.
The pixelmaps are arranged as a 400$\times$280 pixel image, where each pixel corresponds to the reconstructed hits binned by wire number and time ticks respectively, and centered using the reconstructed neutrino interaction vertex. 
The 400$\times$280 pixels represent 400 wires by 1680 time ticks for $\nu_{e}$ events, and 2800 wires by 6720 time ticks for $\nu_{\mu}$ events. 
It is also possible to create 3-D pixelmaps by combining spatial and charge information from all 3 planes.
These 3-D pixelmaps are 100$\times$100$\times$100 pixels while its corresponding true area in the detector is 125$\times$125$\times$250 cm for $\nu_{e}$ events, and 500$\times$500$\times$1000 cm for $\nu_{\mu}$ events
where the last dimension is the direction of the neutrino beam.
These 3-D pixelmaps are used by the CNN in the reconstruction of the 3-D directions, as they directly provide spatial distributions of detector hits.


\begin{figure}[h]
  \vspace{-0.6cm}
  \centering
      \subfloat[Full-event $\nu_{e}$ CC]{\includegraphics[width=.5\textwidth]{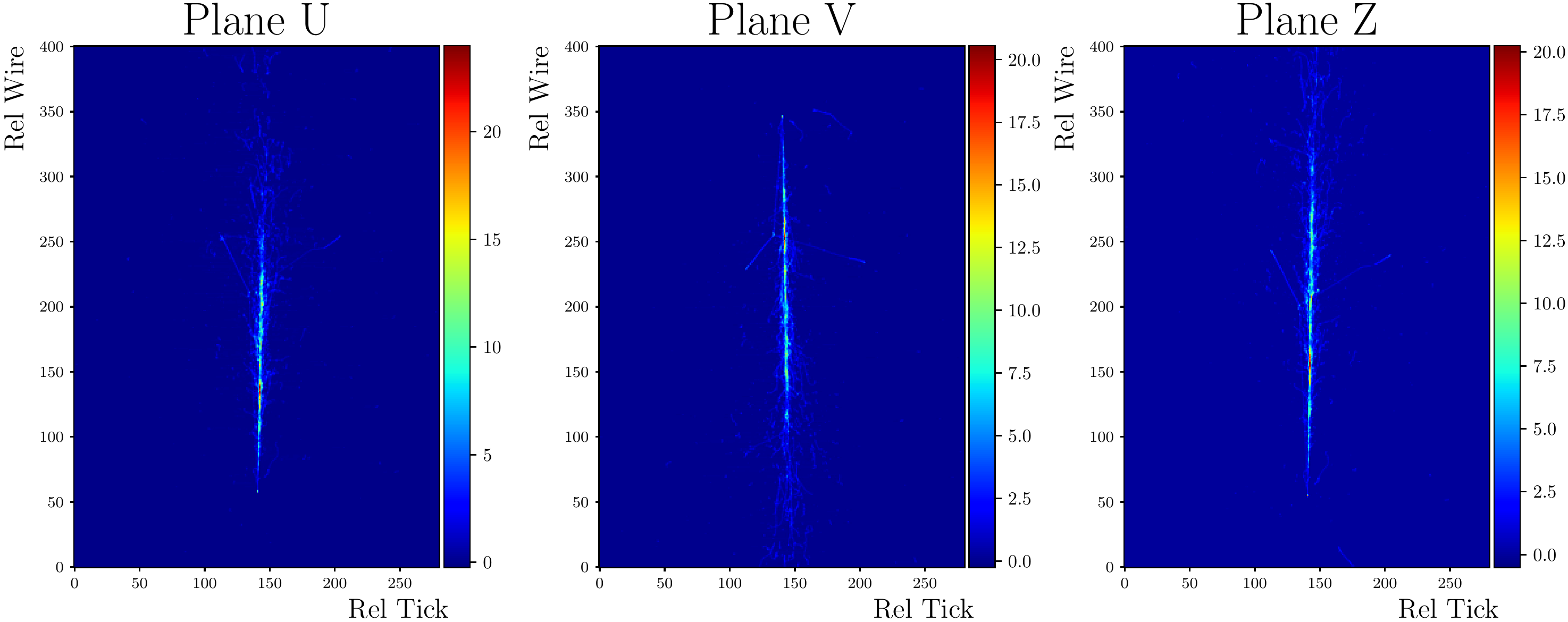}}\hfill
      \subfloat[Full-event $\nu_{\mu}$ CC]{\includegraphics[width=.48\textwidth]{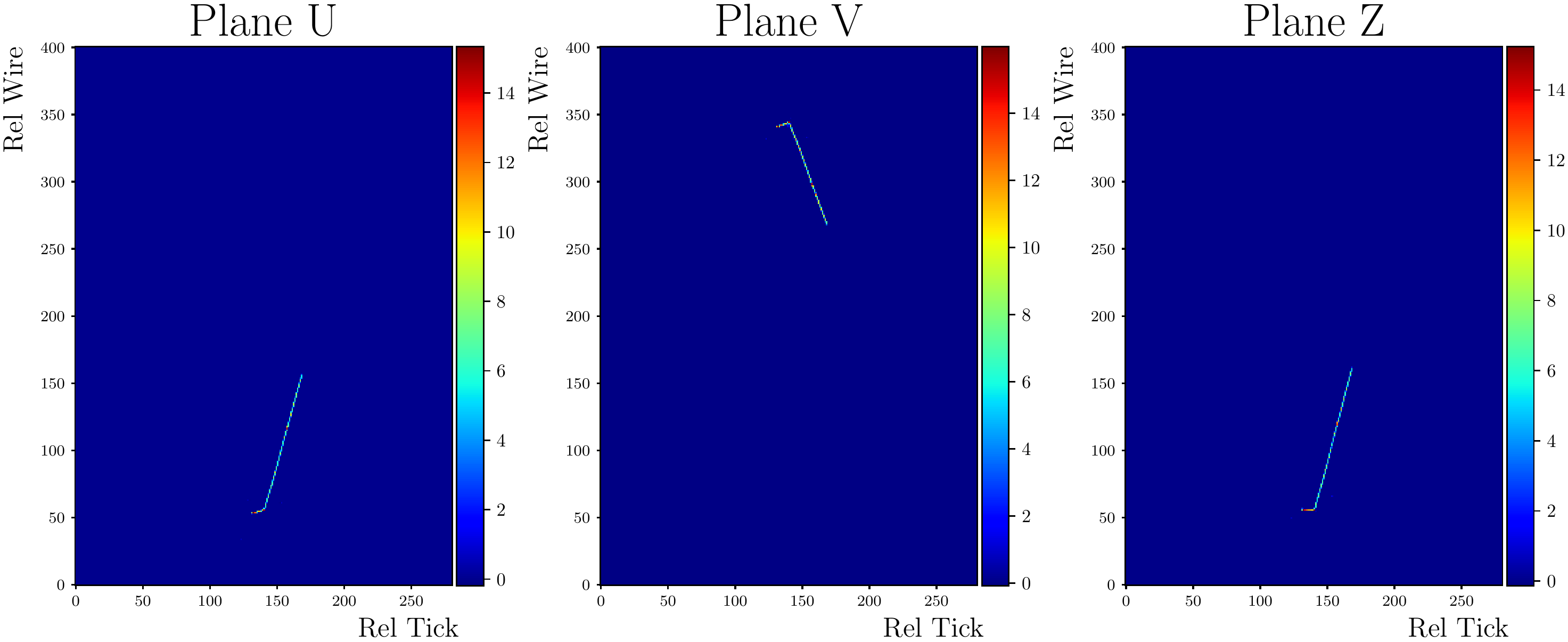}}
  \vspace{-0.1cm}
  \caption{3 views of 2-D pixelmaps for full-event $\nu_{e}$ and $\nu_{\mu}$ CC. 3 views of each event are taken together as inputs of the 2-D energy regression CNN.}
  \vspace{-0.5cm}
  \label{2DPixelMaps}
\end{figure}

\begin{figure}[h]
  \vspace{-0.35cm}
  \centering
  \subfloat[Full-event $\nu_{e}$ CC]{\includegraphics[width=.22\textwidth]{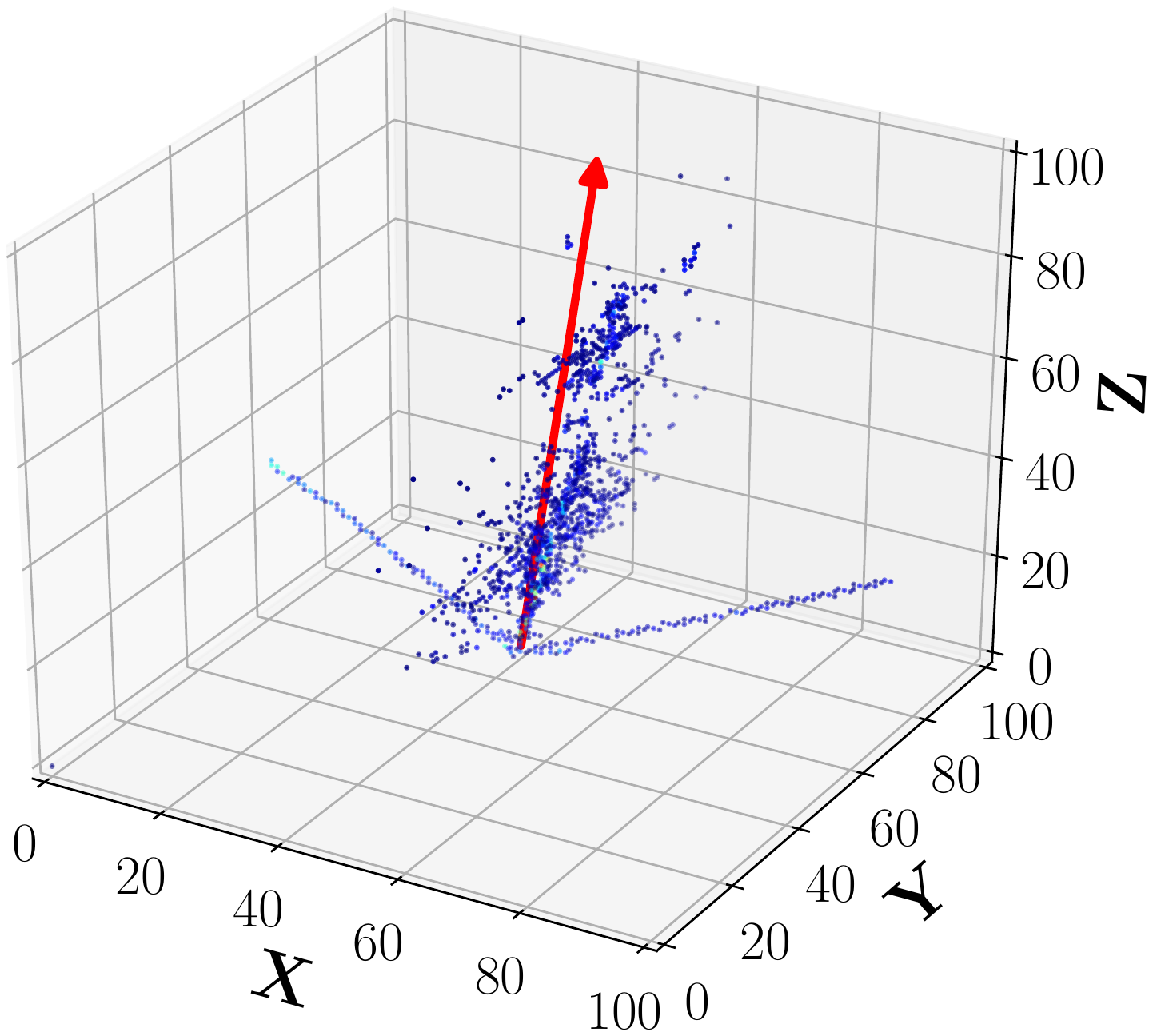}}\hfill
  \subfloat[Prong-only $\nu_{e}$ CC]{\includegraphics[width=.22\textwidth]{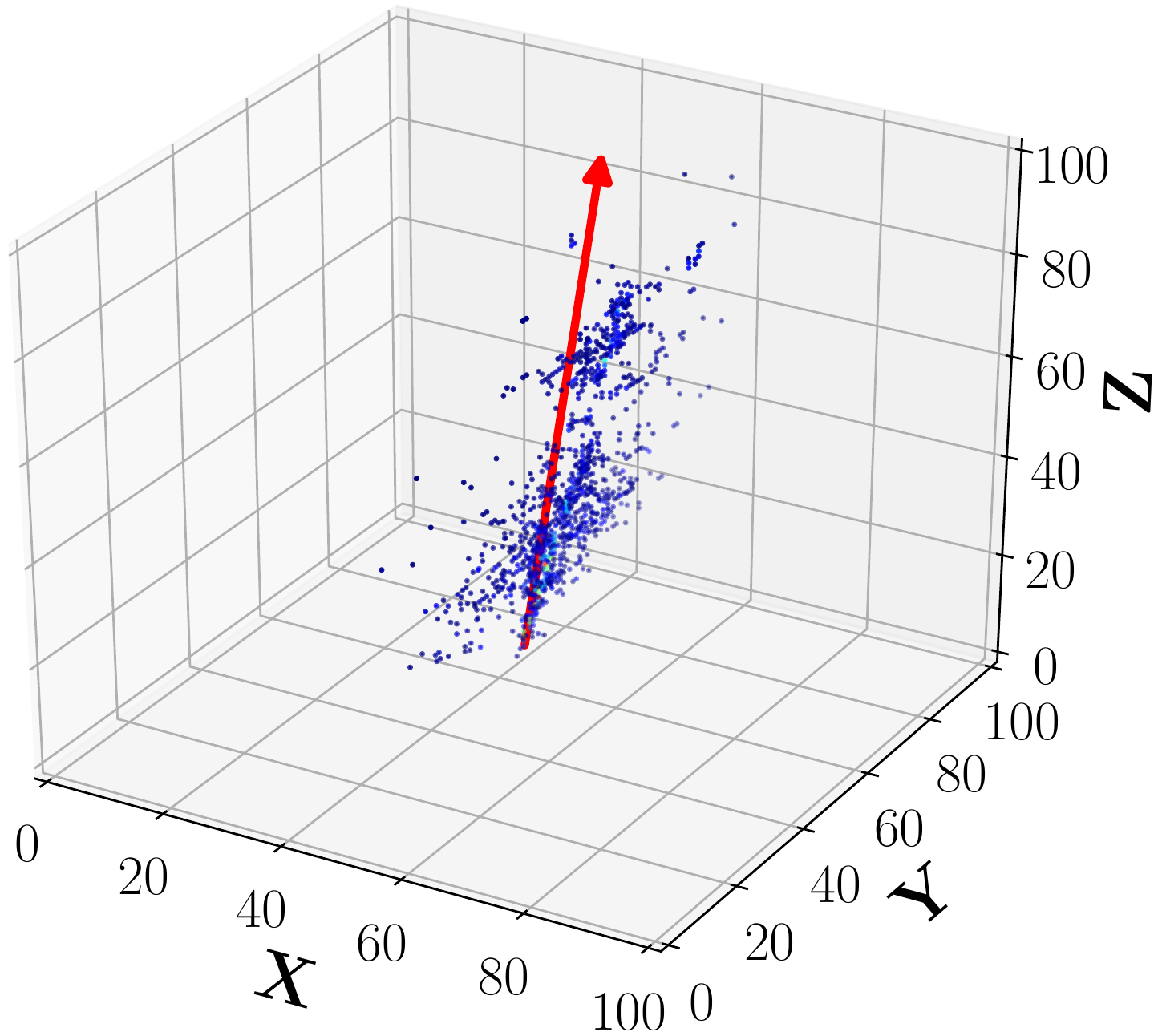}}\hfill
  \subfloat[Full-event $\nu_{\mu}$ CC]{\includegraphics[width=.22\textwidth]{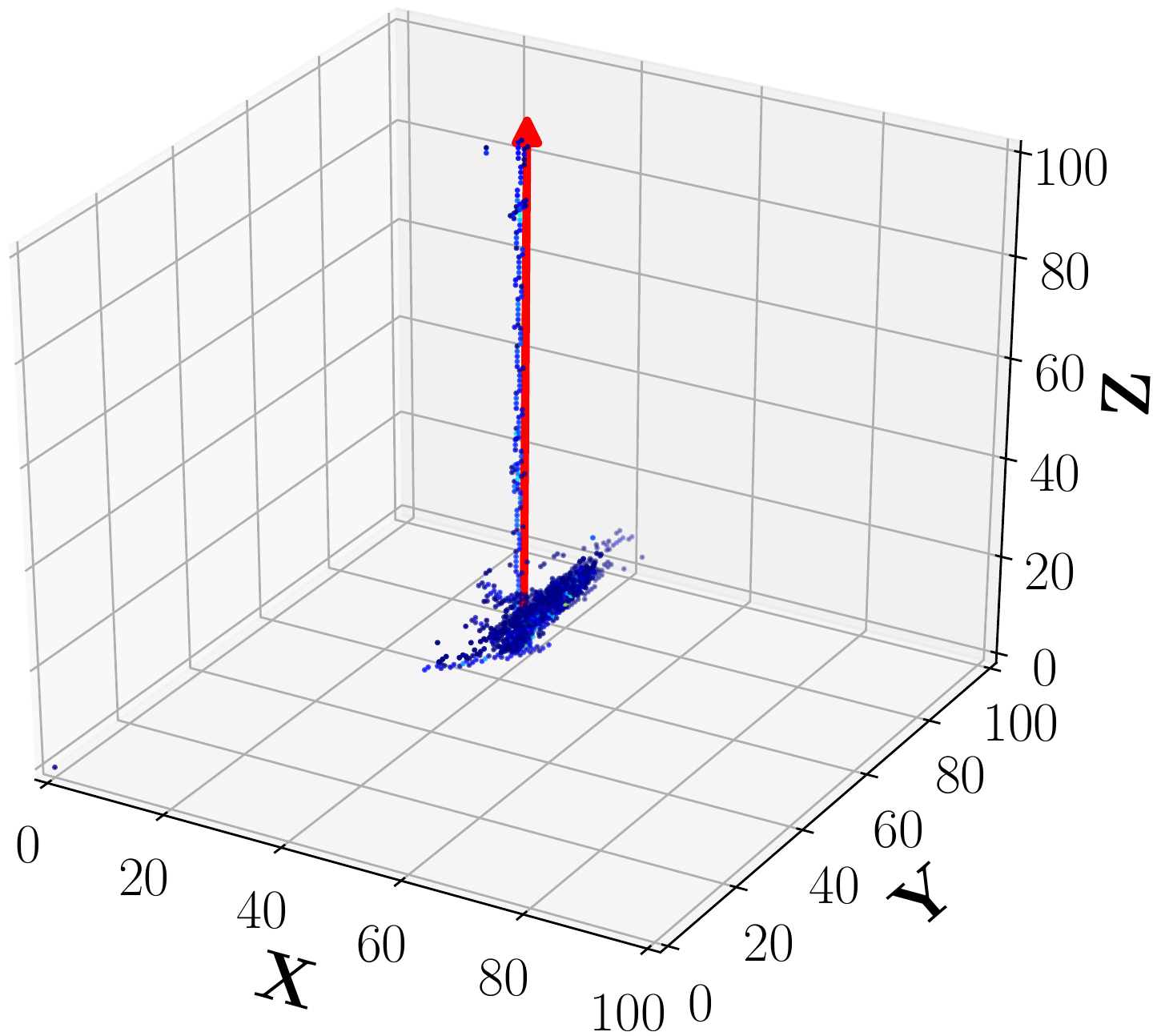}}\hfill
  \subfloat[Prong-only $\nu_{\mu}$ CC]{\includegraphics[width=.22\textwidth]{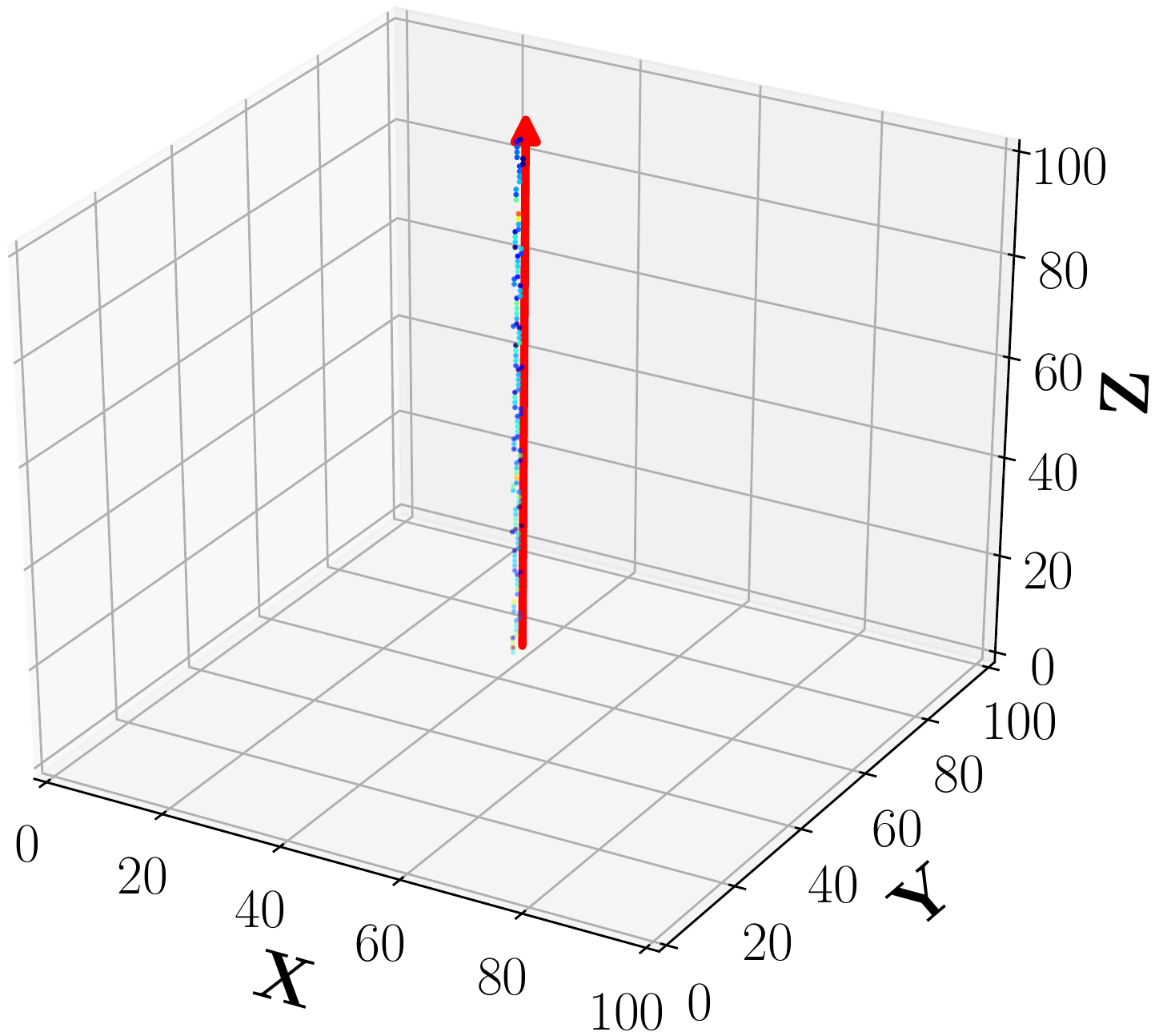}}\hfill
  \vspace{-0.1cm}
  \caption{3-D pixelmaps for full-event and prong-only $\nu_{e}$ and $\nu_{\mu}$ CC. Each pixelmap is taken as an input of the 3-D regression CNN. The red arrows indicate the true directions.}
\end{figure}
\vspace{-0.5cm}

\nsection{Models}
\label{sec:models}

\begin{figure}[t]
\vspace{-0.7cm}
  \centering
  \subfloat[Direction Regression]{\includegraphics[scale=0.8]{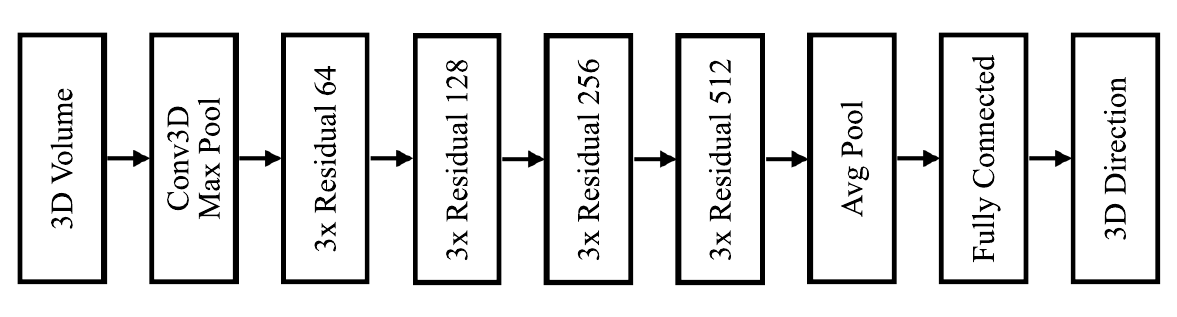}\label{DirectionArchitecture}}
  \hspace{0.05\linewidth}
  \subfloat[Residual block]{\includegraphics[scale=0.43]{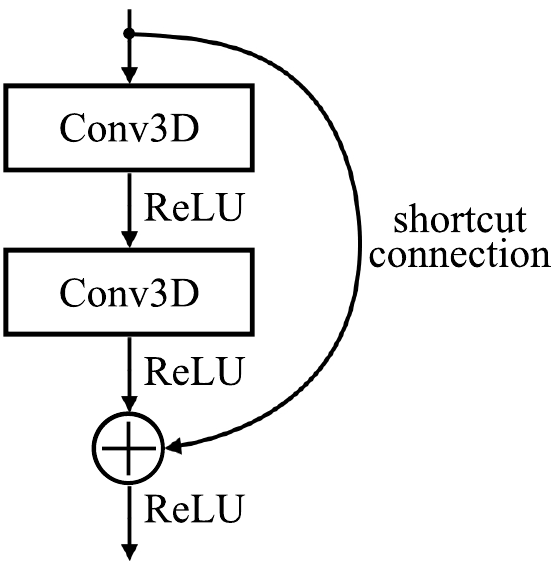}\label{ResidualBlock}}
  \vspace{-0.2cm}
  \caption{Neural network architectures for direction regression.}
  \vspace{-0.7cm}
\end{figure}

We propose two CNN architectures, one for energy reconstruction in 2-D, and another one for direction reconstruction in 3-D. For reconstructing CC and prong energy we use the same architecture as \cite{Seong:DUNE} 
with hyperparameters optimized for these specific tasks using SHERPA \cite{hertel2020sherpa}. 
The inputs to the network are the three plane views U,V, and Z (Figure \ref{2DPixelMaps}). 
For the $\nu_{\mu}$ CC energy the model was trained on a balanced combination of contained and not contained events.
Both $\nu_{\mu}$ CC energy and prong energy models were optimized using Adam \cite{kingma2014adam} with learning rate 0.001, batch size 100 and a mean absolute percentage error loss for up to 100 epochs with early stopping. The 2-D models were trained using Keras \cite{chollet2015keras} with Tensorflow backend \cite{tensorflow2015}.


The direction regression is heavily dependent on the 3-D geometry in the data, thus we designed a 3-D CNN to exploit the structure in the data. 
The model is built on a series of ``residual blocks''  \cite{he2016deep} and a linear layer to output 3-D direction vectors (Figure \ref{DirectionArchitecture}).
Each ``residual block'' includes two convolutional layers with \{64, 128, 256, 512\} number of filters respectively, which are both followed by a batch normalization layer (Figure \ref{ResidualBlock}). 
The input and the output within and between the ``residual blocks'' are connected by the ``shortcut connection''. 
All activation units except the output use Rectified Linear Units (ReLU)~\cite{RectifiedLinearUnits}.
The model is optimized using Adam with learning rate 0.01 with learning rate decay for 200 epochs and mini-batch size 32 in Keras \cite{chollet2015keras}. 
A cosine distance metric was used during the training while a relaxed cosine distance was used for validation and testing. Using regular cosine distance can avoid ambiguity during optimization. It distinguishes between exactly opposite directions, though we can easily infer which hemisphere directions are located in from prior knowledge. Thus we defined relaxed cosine distance loss for better performance as:


 
\vspace{-0.55cm}
\begin{equation}\scriptsize
L_{\mathrm{dir}} = \frac{1}{n}\sum_{i=1}^{n} \min{\left(1 + \frac{\vec{d}^{i}_{\mathrm{True}}\cdot\vec{d}^{i}_{\mathrm{Reco}}}{\left|\vec{d}^{i}_{\mathrm{True}}\right| \left|\vec{d}^{i}_{\mathrm{Reco}}\right|}, 1 - \frac{\vec{d}^{i}_{\mathrm{True}}\cdot\vec{d}^{i}_{\mathrm{Reco}}}{\left|\vec{d}^{i}_{\mathrm{True}}\right| \left|\vec{d}^{i}_{\mathrm{Reco}}\right|}\right)}
\label{eq:cos}
\end{equation}
\vspace{-0.7cm}

\begin{figure}[h]
  \vspace{-0.8cm}
  \centering
  \subfloat[Angular resolution for $\nu_{e}$ CC]{\includegraphics[width=0.44\textwidth]{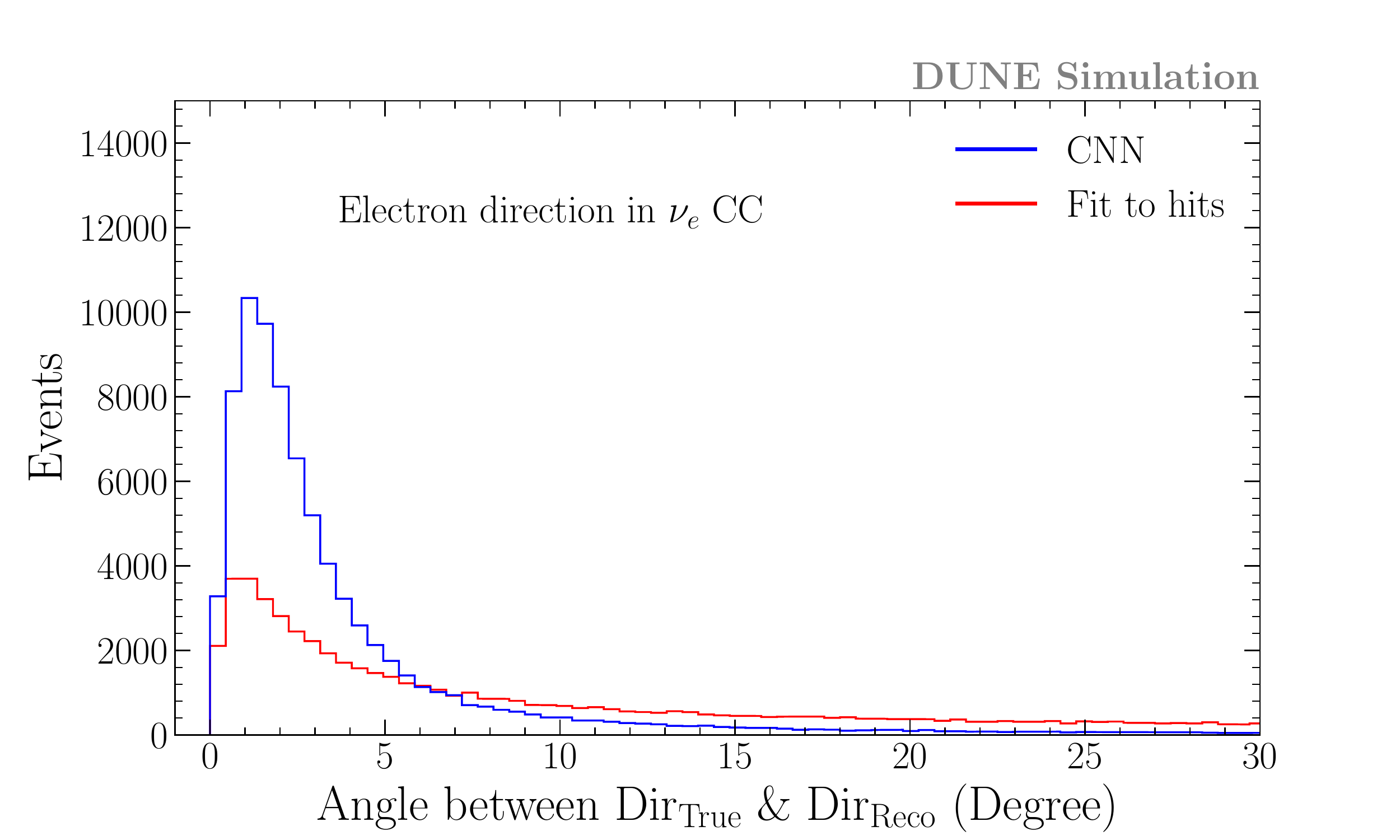}\label{3DNuEAngle}}\hspace{0.1cm}
  \subfloat[Angular resolution for $\nu_{\mu}$]{\includegraphics[width=0.44\textwidth]{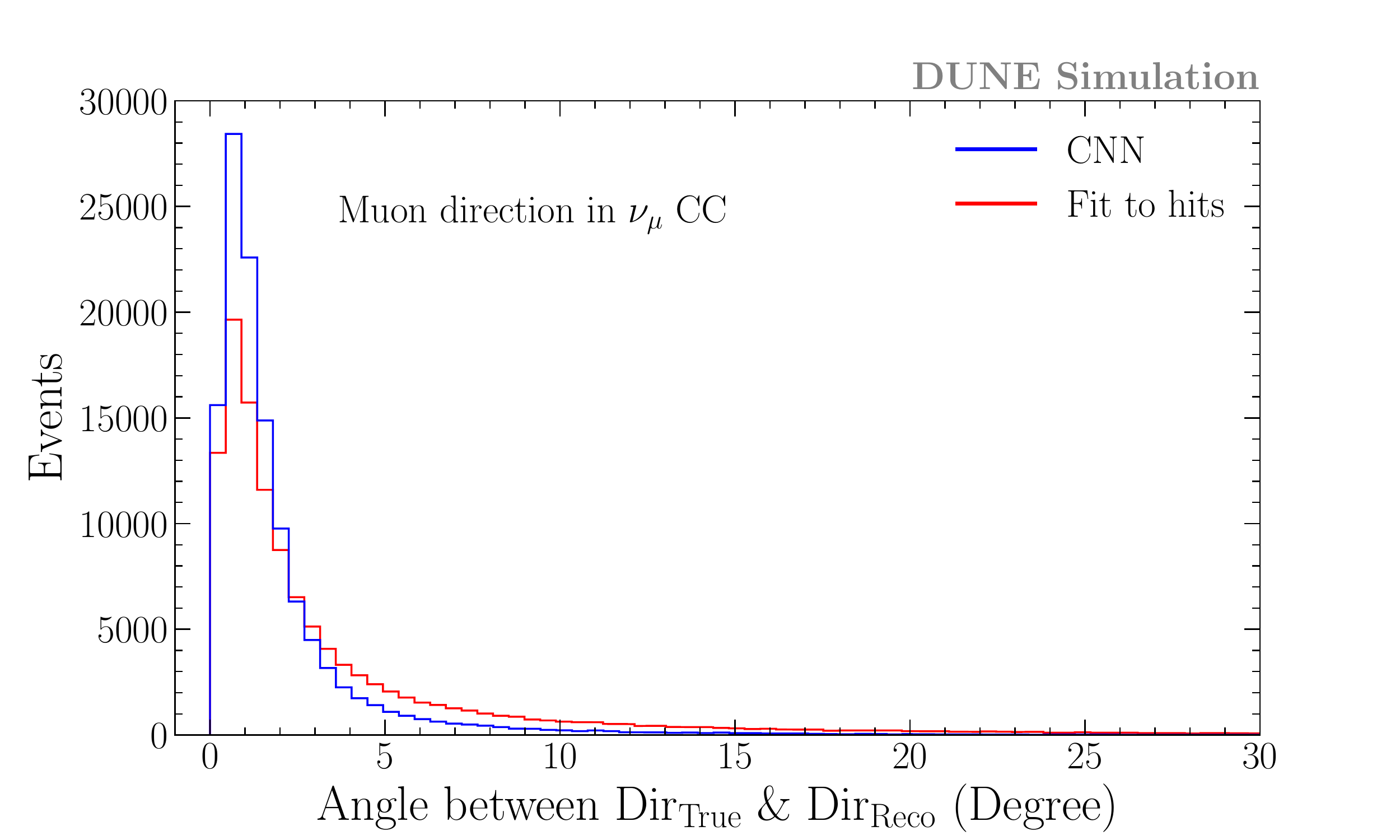}\label{3DNuMuAngle}}
  \\\vspace{-0.43cm}
  \subfloat[Energy dependency for $\nu_{e}$ CC]{\includegraphics[width=0.44\textwidth]{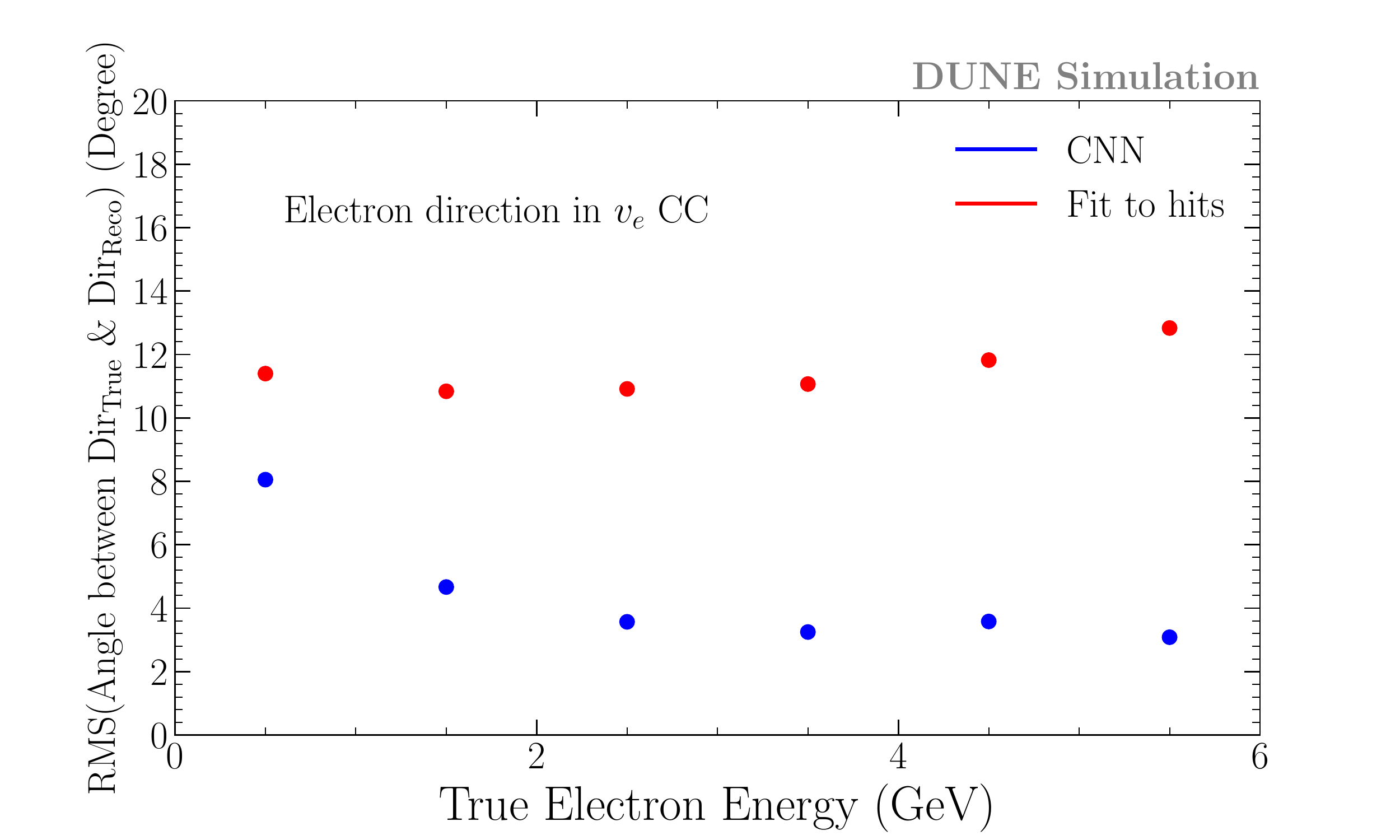}\label{3DNuEEnergyBias}}\hspace{0.1cm}
  \subfloat[Energy dependency for $\nu_{\mu}$]{\includegraphics[width=0.44\textwidth]{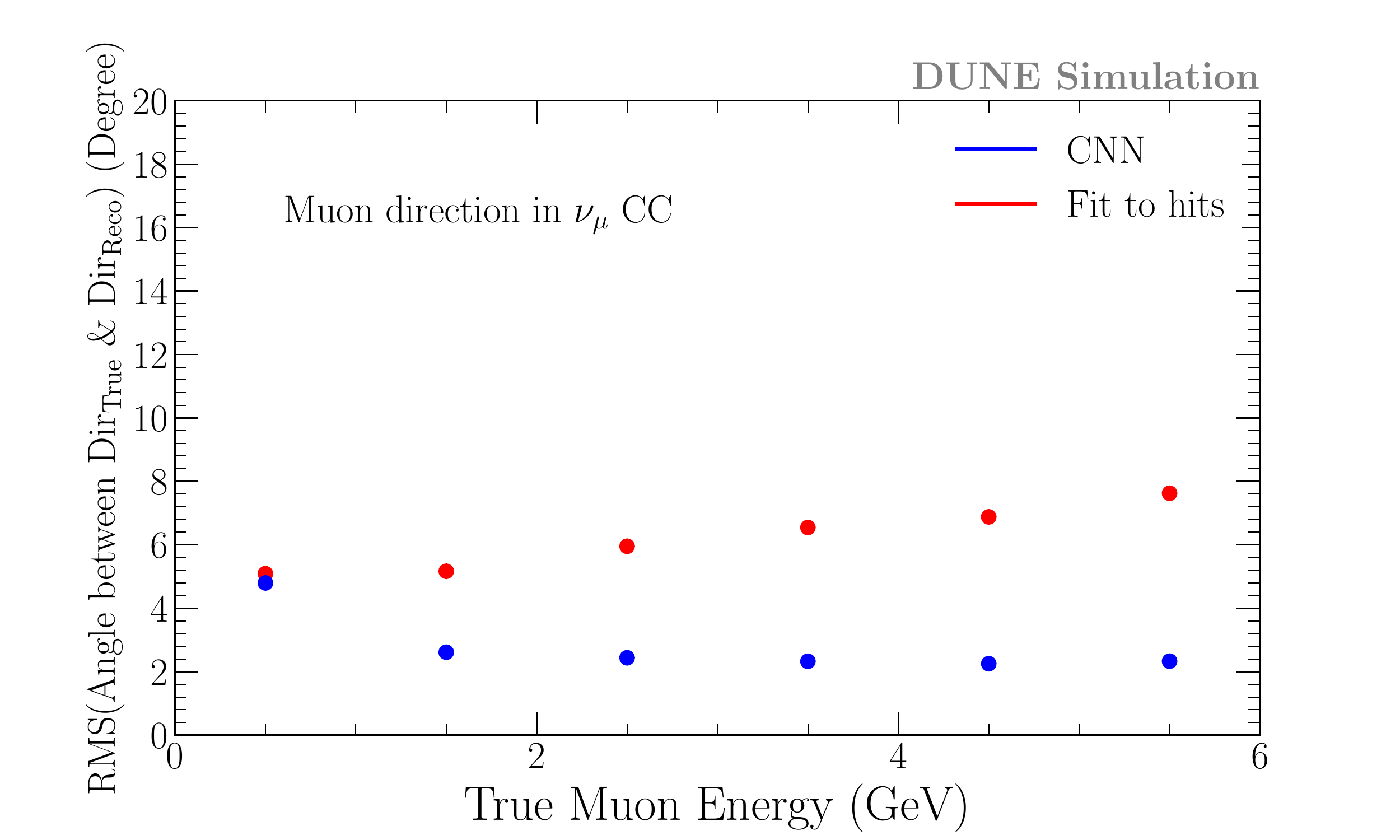}\label{3DNuMuEnergyBias}}
  \\\vspace{-0.1cm}
  \caption{Direction reconstruction performance as angular resolution for $\nu_{e}$ and $\nu_{\mu}$ CC. The 3-D regression CNN produced more precise direction reconstruction than the traditional method.}
  \vspace{-0.3cm}
\end{figure}

\nsection{Results}
\label{sec:results}
Figures \ref{3DNuEAngle} and \ref{3DNuMuAngle} show the distribution of relaxed angular resolutions  
(3-D angle between reconstructed and true directions) 
from the CNN and the traditional method for prong-only $\nu_{e}$ and $\nu_{\mu}$ CC events. 
The angular resolutions are used instead of cosine distances for visualization purposes.
The CNN model produced 13.3$^{\circ}$ and 4.8$^{\circ}$ angular resolution compared with 37.6$^{\circ}$ and 9.5$^{\circ}$ from the traditional method, an improvement of 65\% and 50\% for electron and muon respectively. Figures \ref{3DNuEEnergyBias} and \ref{3DNuMuEnergyBias} show the energy dependence of the RMS of the angular resolutions. The 3-D regression CNN produced a more precise reconstruction for the whole neutrino energy range for both $\nu_{e}$ and $\nu_{\mu}$. These results show the 3-D CNN can extract spatial information better than the traditional clustering and fitting method.


\begin{figure}[h]
  \vspace{-0.8cm}
  \centering
  \subfloat[$\nu_{\mu}$ CC energy (contained).]{\includegraphics[width=0.45\textwidth]{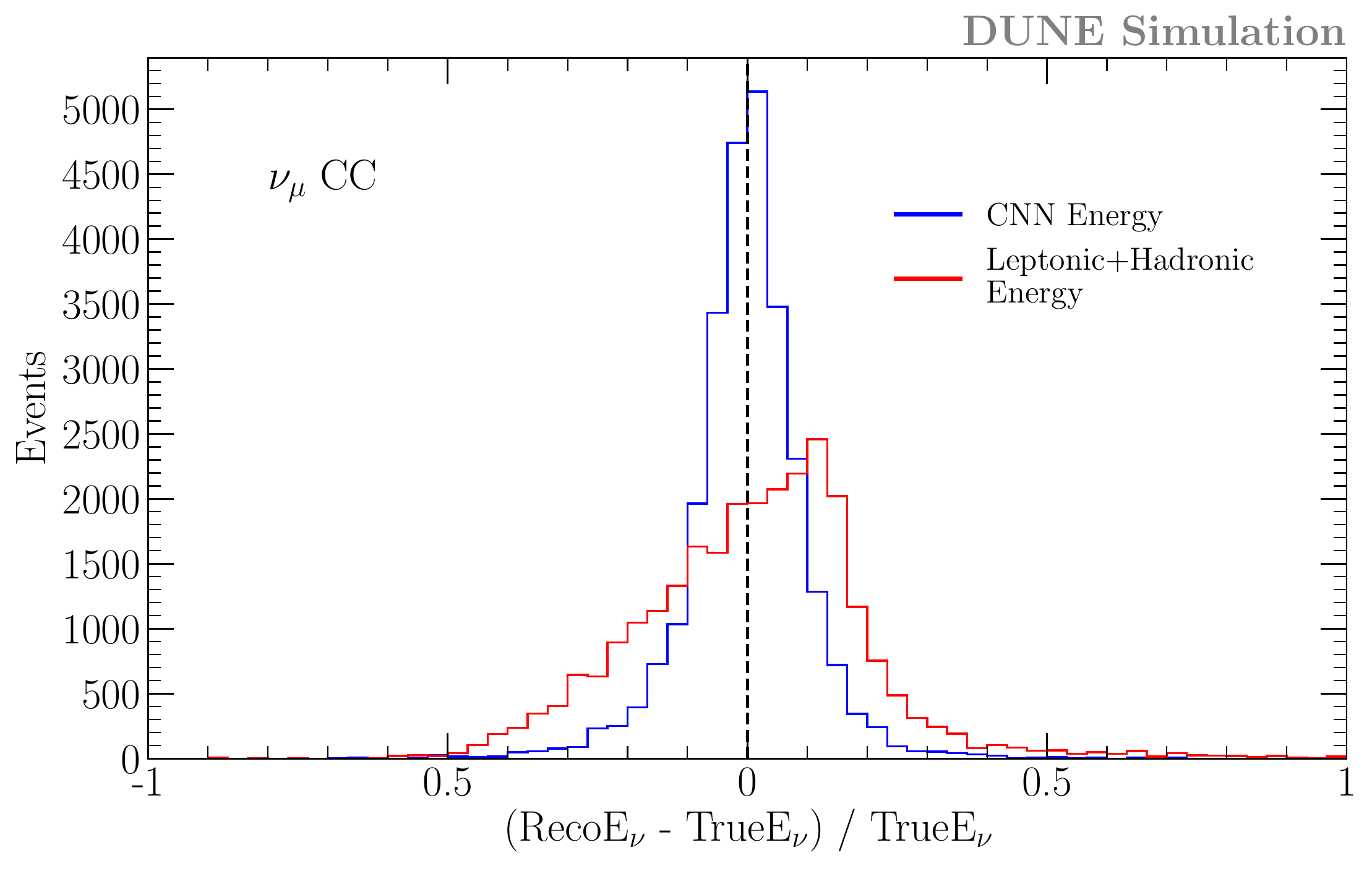}\label{nuMuCCEnergyContained}}\hspace{0.5cm}
  \subfloat[$\nu_{\mu}$ CC energy (not contained).]{\includegraphics[width=0.45\textwidth]{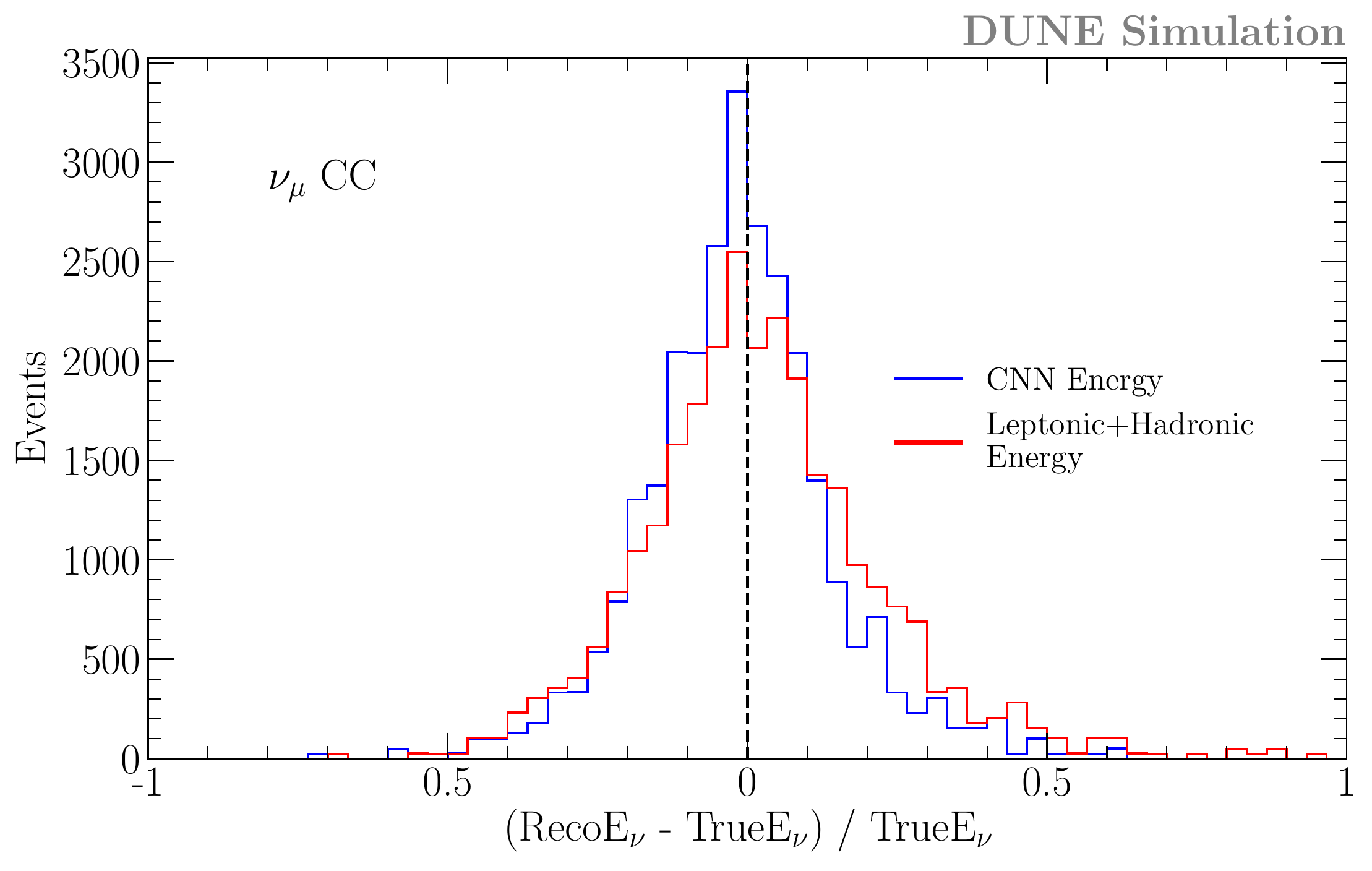}\label{nuMuCCEnergyUncontained}}
  \\\vspace{-0.3cm}
  \subfloat[$\nu_{\mu}$ CC energy (contained).]{\includegraphics[width=0.45\textwidth]{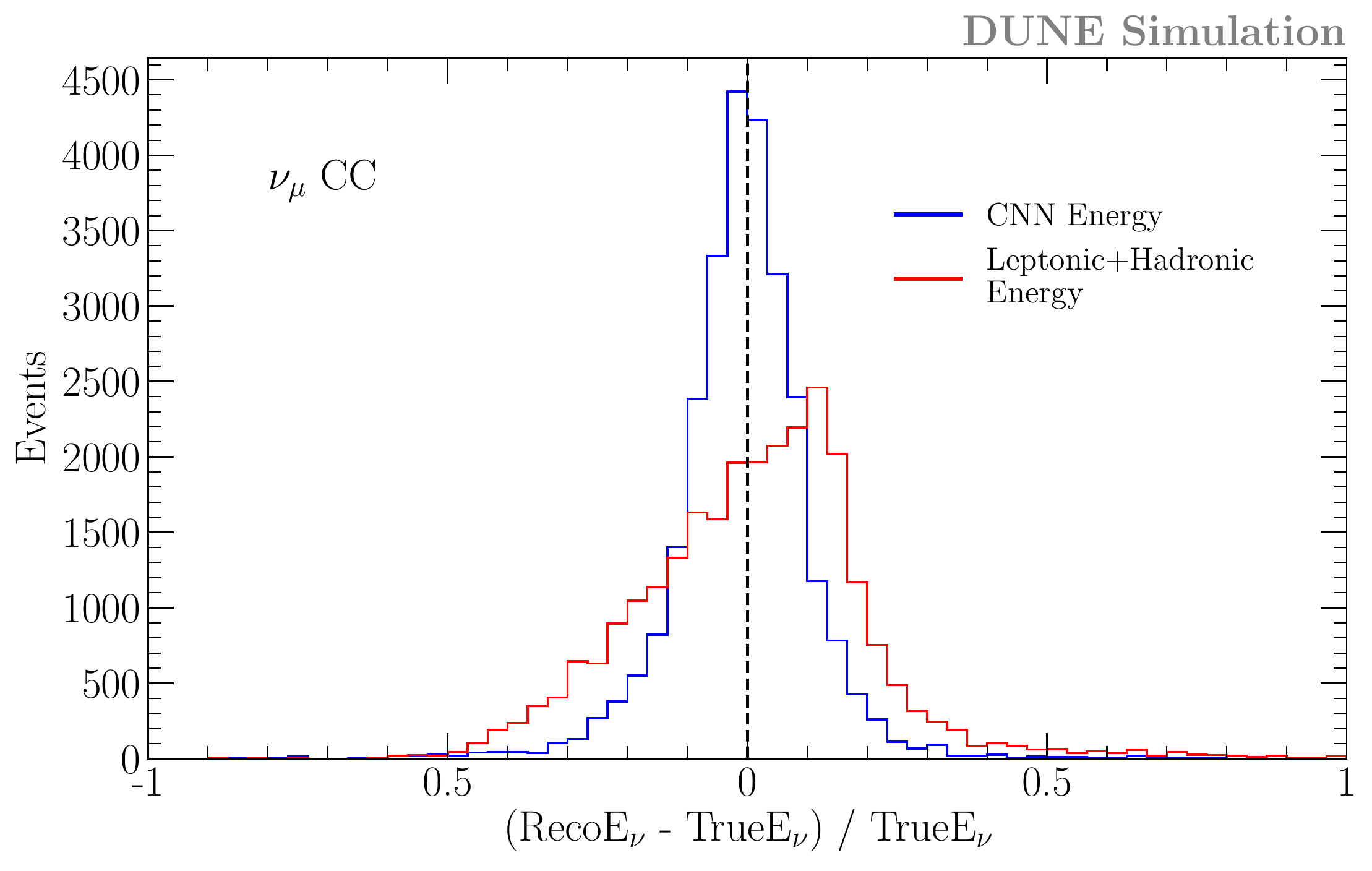}\label{nuMuCCEnergyContainedReweighted}}\hspace{0.5cm}
  \subfloat[$\nu_{\mu}$ CC energy dependency (cont.)]{\includegraphics[width=0.45\textwidth]{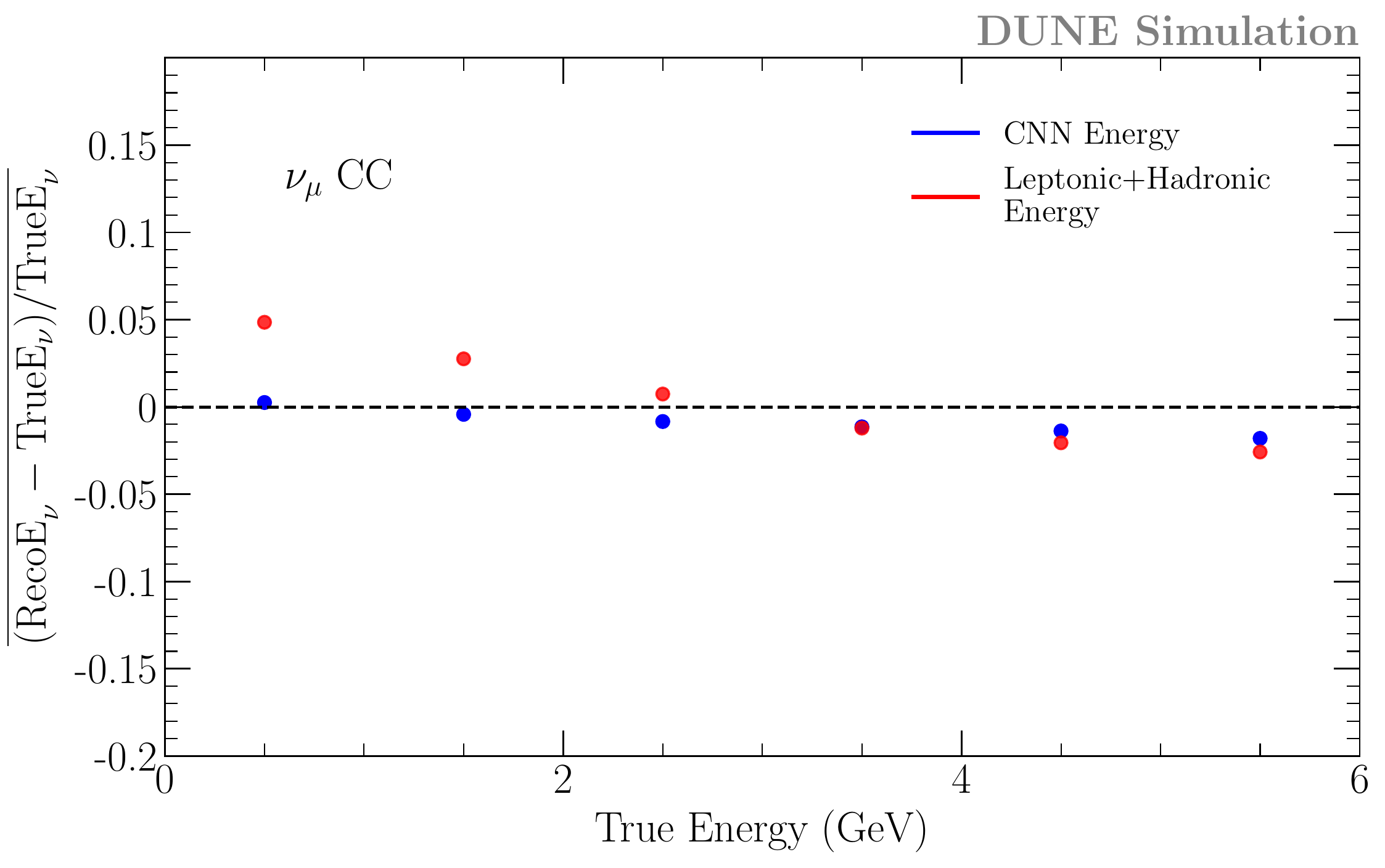}\label{nuMuCCEnergyContainedBias}}
  \\\vspace{-0.3cm}
  \subfloat[Muon prong energy (contained).]{\includegraphics[width=0.45\textwidth]{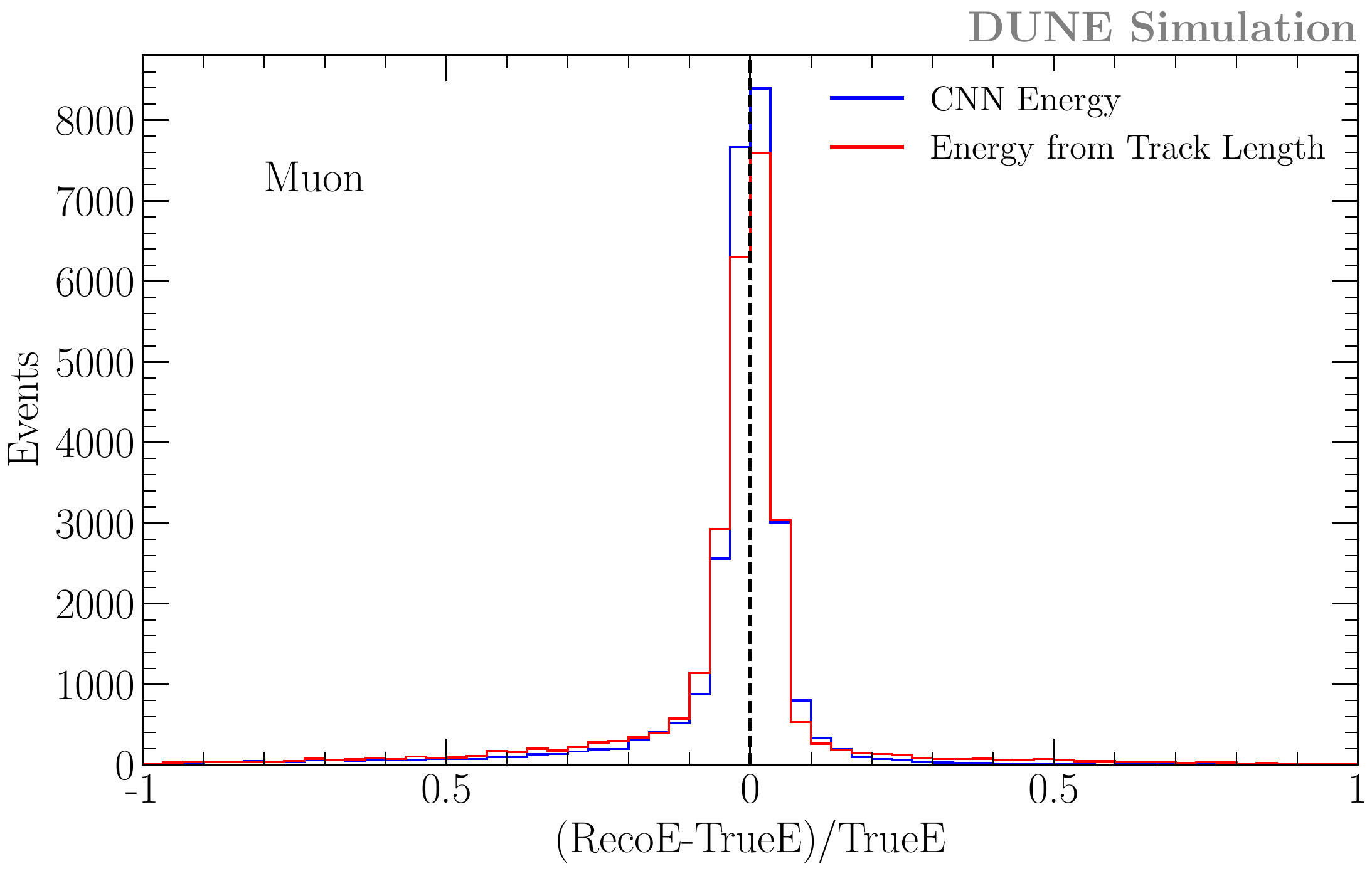}\label{muonEng}}\hspace{0.5cm}
  \subfloat[Electron prong energy.]{\includegraphics[width=0.455\textwidth]{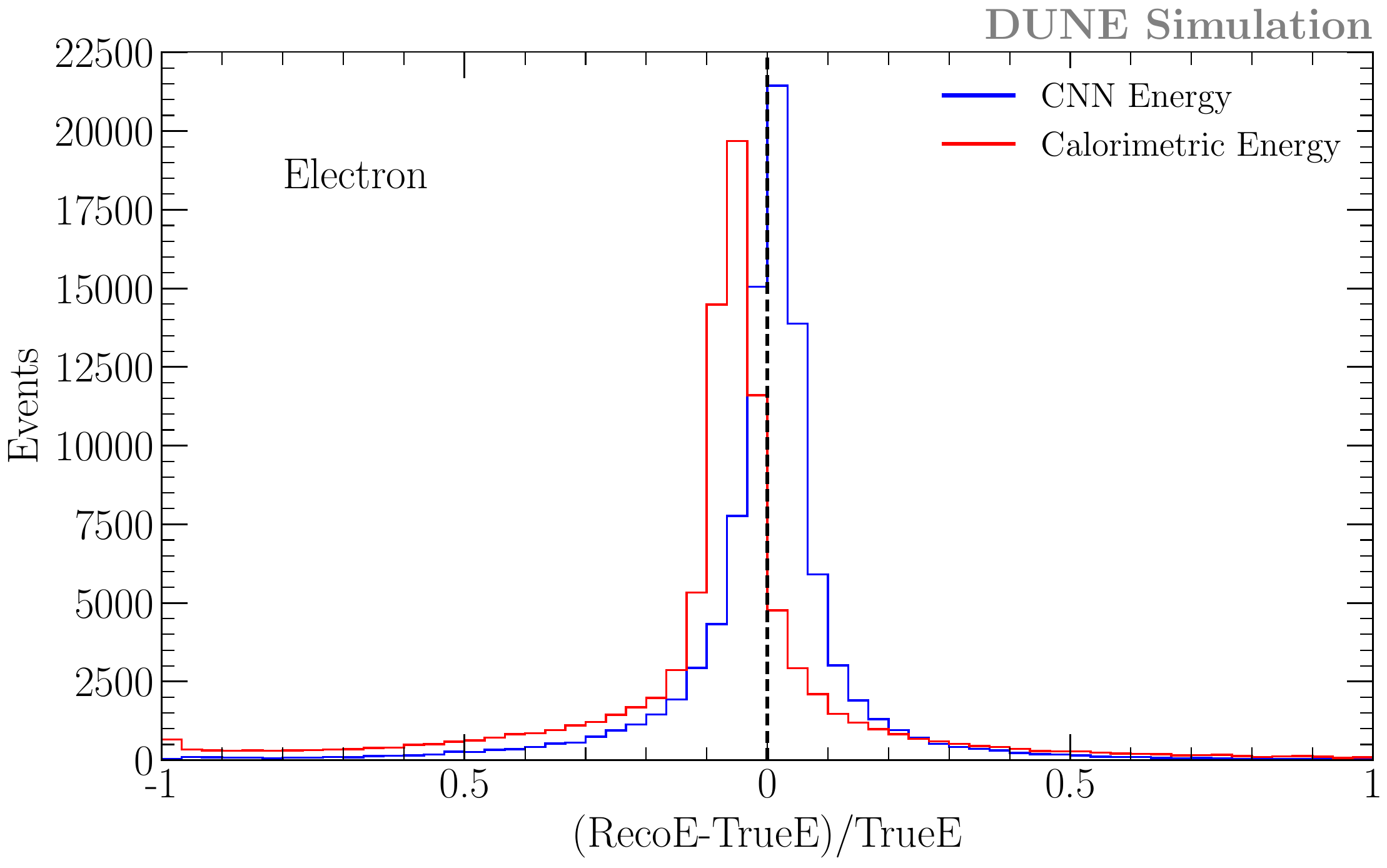}\label{electronEng}}
  \vspace{-0.1cm}
  \caption{$\nu_{\mu}$ CC and final state particle energy reconstruction using CNN.  (c) and (d)  trained with reweighted energy spectrum to address energy dependent biases. The 2-D regression CNN outperformed the traditional method.}
  \label{CCEnergyRegression}
\vspace{-0.2cm}
\end{figure}

Figure \ref{CCEnergyRegression} shows the CNN outperforms the leptonic+hadronic energy method by 47\% and 15\% for $\nu_\mu$ CC events contained and not contained in the detector respectively.  The CNN improves the RMS from 0.191 to 0.101 for contained events and from 0.200 to 0.158 for uncontained events. The improvement for events outside the detector with CNN indicates that AI can automatically and more effectively extract the relationship between the muon scattering angle and uncontained energy without explicitly applying the multi-coulomb scattering equation. Potential energy-dependent biases for contained events were corrected by training CNN with reweighted neutrino energy distribution. The reweighted energy distribution is flat between 0 and 6 GeV and constant for higher energies. However, correcting the bias resulted in a slightly higher resolution (0.116), but it still outperformed the leptonic+hadronic energy method in terms of resolution and  bias (Figure \ref{CCEnergyRegression}). 

Figure \ref{muonEng} and \ref{electronEng} display the track energy reconstruction for the muon (contained events) and electron respectively. The resulting histograms are fit with Gaussian curves to estimate the mean and standard deviation. The CNN and the estimation from the track length perform similar to one another. With both methods achieving a mean of -0.001 and similar resolution around the peak (4\% for muon and 5\% for electron),  while the regression CNN has a much narrower overall distribution, indicating it is less affected by the failure of reconstruction. For the electron, the CNN demonstrates less bias in reconstruction with a mean of 0.007 compared to the Calorimetric Energy of -0.057, an 8 fold improvement in the bias. 

\nsection{Conclusion}
\label{sec:conclusion}
We have developed 2-D and 3-D regression CNNs to reconstruct both spatial and energy  kinematic parameters at DUNE. The regression CNNs outperform the traditional clustering and fitting based methods, indicating that AI can better exploit underlying physics from simulated detector responses. Our models achieve resolution improvements of 65\% for electron directions and 50\% for muon directions, $31$\% for $\nu_\mu$ CC energy, and much smaller RMS for lepton energy. The results are promising for the next phase of DUNE, where high-performance reconstruction algorithms will play an essential role in the analysis of new experimental data. 


\section*{Broader Impact}
\label{sec:broader_impact}
The deep learning algorithms developed in this work could replace many traditional methods in continuous variable reconstruction tasks for complex detectors. This work is a key step towards a full AI-based event reconstruction. By deploying our methods, they will facilitate the analysis of the large volume of experimental data by providing fast and precise kinetic energy and direction reconstruction.

 In the case that our current methods do not match the real data well, detector simulation, energy calibration, and neutrino event generator can be tuned based on the observed difference between data and simulation. Data-driven models and training could also be implemented in our algorithms to mitigate this issue. While our models are currently trained on DUNE simulation, the next step is to validate their performance on the data taken by DUNE's prototype detectors at CERN. 


\bibliographystyle{JHEP}
\bibliography{neurips_2020.bib}

\end{document}